# SmallSat Solar Axion and Activity X-ray Imager (*SSAXI*)

Jaesub Hong[1], Suzanne Romaine[2], Almus Kenter[2], Christopher Moore[2], Katherine Reeves[2], Brian Ramsey[3], Kiranmayee Kilaru[3], Julia K. Vogel[4], Jamie Ruz Armendariz[4], Hugh Hudson[5], Kerstin Perez[6]

[1] Harvard Smithsonian Center for Astrophysics, Cambridge, MA 02138

[2] Smithsonian Astrophysical Observatory, Cambridge, MA 02138

[3] NASA Marshall Space Flight Center, Huntsville, AL 35808

[4] Lawrence Livermore National Laboratory, Livermore, CA 94550

[5] University of California, Berkeley, CA 94720

[6] Massachusetts Institute of Technology, Cambridge, MA 02139

## ABSTRACT

Axions are a promising dark matter candidate as well as a solution to the strong charge-parity (CP) problem in quantum chromodynamics (QCD). We describe a new mission concept for SmallSat Solar Axion and Activity X-ray Imager (*SSAXI*) to search for solar axions or axion-like particles (ALPs) and to monitor solar activity of the entire solar disc over a wide dynamic range. *SSAXI* aims to unambiguously identify X-rays converted from axions in the solar magnetic field along the line of sight to the solar core, effectively imaging the solar core. *SSAXI* also plans to establish a statistical database of X-ray activities from Active Regions, microflares, and Quiet Sun regions to understand the origin of the solar corona heating processes. *SSAXI* employs Miniature lightweight Wolter-I focusing X-ray optics (MiXO) and monolithic CMOS X-ray sensors in a compact package. The wide energy range (0.5 – 6 keV) of *SSAXI* can easily distinguish spectra of axion-converted X-rays from typical X-ray spectra of solar activities, while encompassing the prime energy band (3 – 4.5 keV) of axion-converted X-rays. The high angular resolution (30 arcsec HPD) and large field of view (40 arcmin) in *SSAXI* will easily resolve the enhanced X-ray flux over the 3 arcmin wide solar core while fully covering the X-ray activity over the entire solar disc. The fast readout in the inherently radiation tolerant CMOS X-ray sensors enables high resolution spectroscopy with a wide dynamic range in a broad range of operational temperatures. *SSAXI* will operate in a Sun-synchronous orbit for 1 yr preferably near a solar minimum to accumulate sufficient X-ray photon statistics.

## 1. INTRODUCTION

SmallSats and CubeSats are becoming increasingly popular platforms to conduct leading science investigations at low cost. They enable rapid technology developments through quick turnaround cycles. SmallSat Solar Axion and Activity X-ray Telescope (*SSAXI*) is a new SmallSat science mission concept designed to search X-ray signatures of solar axions and to monitor solar X-ray activities of Active Regions (ARs), microflares, and Quiet Sun (QS) regions. The Science Objectives of *SSAXI* ambitiously address some long-standing questions in Astrophysics and Heliophysics – the nature of dark matter and the origin of the coronal heating processes. The main instrument in *SSAXI* takes advantage of some of the recent advances in space X-ray instrumentation such as Miniature lightweight X-ray optics [1] and CMOS X-ray sensors [2]. Collaborating with mature commercial CubeSat/SmallSat spacecraft vendors that utilize cost-efficient commercial off-the-shelf components, we envision that *SSAXI* can be quickly developed at low cost in ~ 3 –

---

[1] jaesub@head.cfa.harvard.edu; phone 1 617 496-7512; fax 1 617 496-7577

4 years. In **Sections 2** and **3**, we briefly review the two main science objectives of *SSAXI*. In **Sections 4** and **5**, we outline the instrument design and the overall mission concept.

## 2. AXIONS AND SEARCH FOR SOLAR AXIONS

The nature of dark matter remains one of the fundamental mysteries in astrophysics and cosmology. Currently the leading candidates for dark matter are Weakly Interacting Massive Particles (WIMPs) and axions. Originally postulated by Peccei and Quinn, the axion is a hypothetical elementary particle arising from the most viable solution for the strong charge-parity (CP) problem in quantum chromodynamics (QCD) of the standard model [3]. The strong CP problem is the puzzling, apparent conservation of the combined symmetry of charge+parity in strong interactions, although the mathematical formulation of the QCD does not prohibit a violation of the CP symmetry.

Standard axions of a symmetry-breaking scale (of the order of the electroweak interaction) were quickly ruled out [4, 5], but newer models of arbitrary scales were developed by Kim-Shifman- Vainshtein-Zakharov (KSVZ) [6, 7] and Dine-Fischler-Srednicki-Zhitnitskii (DFSZ) [8, 9]. These lighter axions, and the more general Axion-Like Particles (ALPs), which are well motivated by string theory, are postulated to interact so weakly that they are called "invisible". Nevertheless, these theoretically inspired axions or ALPs would have far-reaching consequences in astrophysics and cosmology. For instance, ALPs, which are expected to be generated in the hot thermal plasma of stellar cores, provide a new process of energy loss in stellar evolution [10]. According to inflation theory, primordial axions of low kinetic energy should have been created abundantly [11]. These primordial axions of low mass (< ~1 meV) are a particularly attractive candidate for dark matter because many would have survived and filled the universe because of the lack of a decay process into lighter particles [12].

Of the many observational techniques that have been developed over the years in search of axions, perhaps one of the most attractive is to look for axions emitted from the core of the Sun. Axions or ALPs are expected to emerge abundantly from the hot plasma in a stellar core through the incoherent Primakoff effect, where a photon converts into an axion in the electric field of a charged particle (**Fig. 1A**). These axions or ALPs provide a new process of energy loss in stellar evolution [10]. The emerging axions would then have a blackbody thermal distribution representative of the solar interior with peak and mean energies of roughly 3 and 4 keV, respectively (these values might be shifted to lower energies for some axion models where axions directly couple to electrons). Many axion search/detection methods rely on the inverse *coherent* Primakoff effect, where an ALP, otherwise invisible, is re-converted to an X-ray photon by a transverse magnetic field. In the case of solar axions, this conversion can occur in the magnetic field of the solar atmosphere or in the magnetic field of a laboratory (e.g. *CAST* [13]).

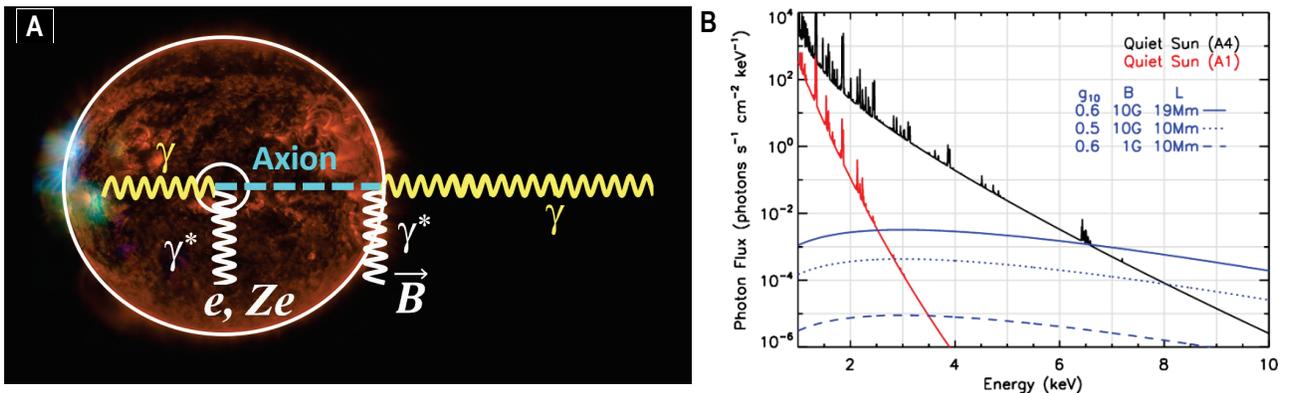

**Fig. 1 (A) Axion creation in the solar core by the Primakoff effect and its conversion to an X-ray in the solar magnetic field by the inverse Primakoff effect. (B) Solar X-ray fluxes during two quiet sun states (black & red; scaled for a 0.1 R$_\odot$ disc region) [14] in comparison with the expected axion-converted X-ray signal (blue) from the central 0.1 R$_\odot$ disc [15].**

**Fig. 1B** compares the solar X-ray spectra during a quiet sun state (A4: ~4 ×10$^{-8}$ W m$^{-2}$) and a "deep solar minimum" state (<A1: ~7×10$^{-10}$ W m$^{-2}$) with the expected axion-converted X-ray fluxes from the central 0.1 $R_\odot$ disc (~50% of the total flux). The expected number of photons is proportional to $g^4 (B L)^2$ where $g$, $B$ and $L$ are the axion-photon coupling constant, the magnetic field strength and conversion length, respectively. Given the size of the solar core and the expected axion-converted X-ray spectra, *the axion or ALP-induced X-ray signature can be determined decisively when it is simultaneously spatially resolved and spectrally detected* with a sensitive soft X-ray imaging spectrometer; the imaging or spectral signature by itself would be inconclusive and would allow other interpretations. A successful detection would not only be of cosmological significance, but it would also provide a unique opportunity to observe the core of the Sun.

**Fig. 2** compares the exclusion region in the coupling ($g$) versus axion mass space set by leading axion search experiments with the upcoming *NuSTAR* observations and one Compact X-ray Imaging Spectrometer (CXIS) module of *SSAXI*. Even with a relatively small effective area (>~2 cm$^2$), a dedicated 1 yr observation with CXIS could outperform a ~ 100 ks *NuSTAR* observation by a factor of ~10 in terms of axion signal detection due to a lower background (<~0.01×; mainly due to a large unocculted opening (~2 – 5 deg) of the *NuSTAR* optics [16]) and the fast readout system in CXIS (cf. *NuSTAR* will have ~75% dead time due to high background solar X-ray fluxes). The CXIS fast readout system (**Section 4**) also enables effective observations for a wider range of solar states, whereas *NuSTAR* axion search would be limited to sub-A solar states.

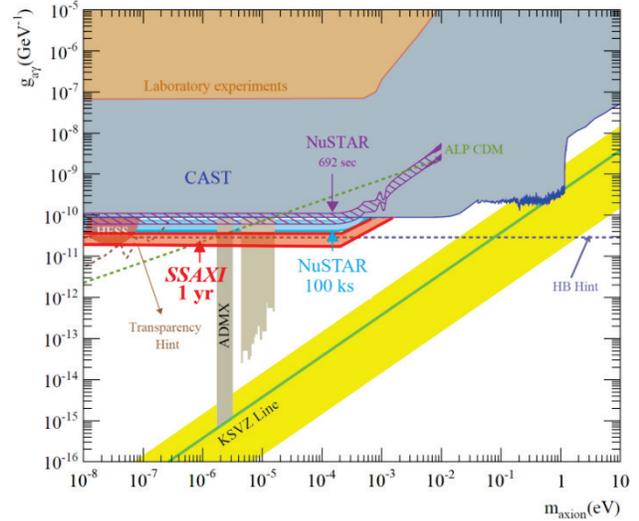

**Fig. 2** Exclusion region in the axion-photon coupling ($g_{a\gamma}$) versus the axion rest mass ($m_a$) set by various experiments overlaid with the expected performance (preliminary) of a 100 ks *NuSTAR* observation (cyan; 75% dead time) and a 1yr *SSAXI* observation near a solar minimum (red; 70% duty cycle).

Given the rapid development using SmallSat platforms, *SSAXI* is not only complementary but also competitive relative to next generation axion search programs like *IAXO* [17, 18] or *ALPS* II [19], which are at least a decade away. In comparison to laboratory experiments such as *ADMX* [20] or *ABRACADABRA* [21], helioscope searches like *SSAXI* are indifferent to assumption of the dark matter density in the universe.

### 3. UNDERSTANDING SOLAR CORONAL HEATING FROM MICROFLARES AND QUIET REGIONS IN COMPARISON WITH ACTIVE REGIONS

Solar flares are large, flashy events that release a substantial amount of energy, but they are not frequent enough to be responsible for heating the solar corona. Smaller events, often referred to as "microflares" (defined here as *GOES* C class and below), happen more frequently, but release a smaller amount of energy. It has been postulated that even smaller, but ubiquitous reconnection events ("nanoflares") could release enough energy to heat the corona [22, 23]. In order to flesh out this possibility, it is necessary to understand flare energy release on small scales.

Large flares occur in active regions and often have complicated magnetic field structures, while microflares are small events that can occur in simpler field geometries that could easily be more common in solar environments. Thus, it is not necessarily clear that the triggering mechanism and subsequent energy release will be the same in both large flares and microflares, and small reconnection events could be far more numerous than has been previously assumed. The energy released in large flares has been found to mostly go into accelerating electrons, with smaller but roughly equal contributions to kinetic energy in the coronal mass ejection (CME) and direct heating of the plasma [24]. It is possible that the energy partition in microflares is different from that in large flares. For example, small flaring events are less likely to involve CME-like eruptions, so there may be more energy channeled into direct heating of the plasma rather than into the kinetic energy of a plasmoid as in a CME.

Some previous statistical studies carried out with *RHESSI* on small flares (between *GOES* A and C class) indicate that the frequency of microflares with respect to their thermal energy output is similar to large flares [25], and that microflares have similar correlations between various thermal and non-thermal parameters as large flares [26]. However, these correlations should be viewed with caution, since there is a selection bias in *RHESSI* observations that favors hotter events due to its low sensitivity to soft X-rays below ~3 keV [26].

There is no statistical knowledge of the thermal properties of the smallest flares, given the limited spectral measurements below the *GOES* A level. The astronomy mission *NuSTAR* occasionally carries out solar observations, and during one of these special observations Glesener et al. (2017) observed a small sub-A level flare that seemed to have similar properties to larger flares [27]. Wright et al. (2017) also observed a microflare with *NuSTAR*, and found that the spectrum was purely thermal, with emission extending up to 10 MK in the impulsive phase [28]. A few microflares have also been observed with the *FOXSI* rocket experiments [29, 30]. Some microflares below *GOES* A-level were detected near the last solar minimum by the SphinX spectrometer on the Russian *CORONAS-Photon* mission [31, 32], but these observations are spatially and spectrally integrated. The *MinXSS* Cubesats [33] have spectrally resolved spectra from 0.8 – 12 keV, but do not have spatial information, making the study of microflares during solar maximum challenging. Thus, there have been only a handful of solar spatially separated, spectrally resolved measurements of microflares in the spectral range of 0.5 – ~4 keV. The *SSAXI* mission aims to cover this missing phase space through spatially (<~30 arcsec HPD), spectrally (<~120 eV FWHM) and temporally (<0.025 sec) resolved measurements of microflares in the spectral range of ~0.5 – 5 keV over the entire solar disc. Fig. 3 illustrates example data cubes of *SSAXI* observations.

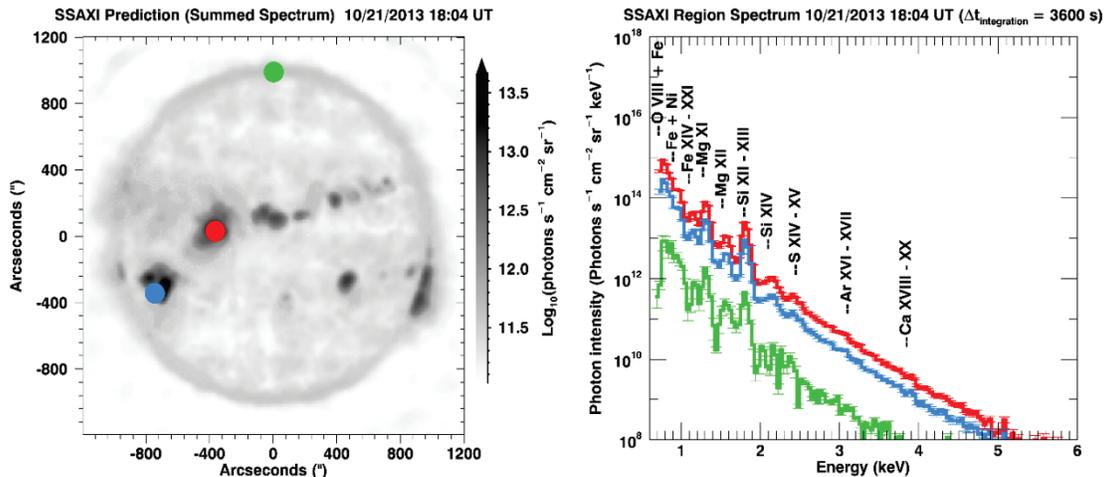

**Fig. 3 Predicted *SSAXI* signal calculated from the Sparse DEM (from *SDO*/AIA UV images). (Left) Spectrum integrated signal map with 3 example regions indicated by red (AR), blue (microflare) and green (QS) dots. (Right) Spectral intensity from the 3 respective regions. *SSAXI* spectral data cubes will examine the spatial variability of the coronal temperature structure and elemental abundances of Fe, Mg, Si, S, Ar and Ca. There is large variability between the spectra (from temperature and abundances) that is hidden in spectrally integrated images similar to *Hinode*/XRT and *Yohkoh*/SXT, and spatially integrated spectra similar to *MinXSS*/X123, *MESSENGER*/SAX and *CORONAS*/SphinX. *SSAXI* will perform new measurements that have eluded solar physics for years.**

# 4. COMPACT X-RAY IMAGING SPECTROMETER (CXIS)

*SSAXI* employs a set of Compact X-ray Imaging Spectrometers (CXIS) to achieve the Science Objectives described in the previous sections. The CXIS is a lightweight Wolter-I X-ray telescope in a small form factor, consisting of a Miniature X-ray Optics (MiXO) module and a monolithic CMOS X-ray sensor as illustrated in **Fig. 4**. A common backend electronics module with a Controller Board (CTB) powers and controls multiple focal plane CMOS detectors, and processes their data. **Table 1** summarize key instrument parameters of one CXIS module.

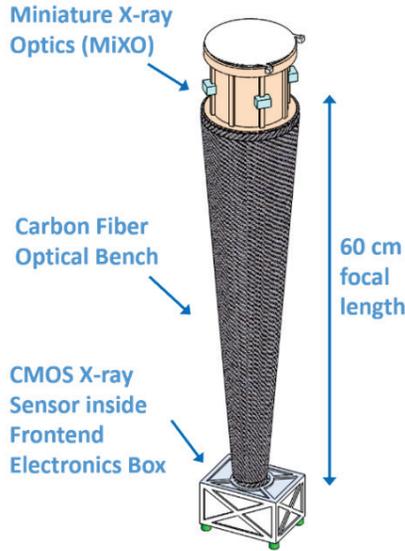

**Fig. 4** Compact X-ray Imaging Spectrometer consisting of Miniature lightweight Wolter-I X-ray Optics (MiXO) and a monolithic CMOS X-ray sensor.

**Table 1** Key Instrument parameters of one CXIS module in *SSAXI*

| Parameters | One CXIS Module |
|---|---|
| Main Subsystems | MiXO + CMOS + backend electronics |
| Volume<br>Mass | 15×15×70 cm<br>7 kg |
| Power | ~6 – 7 W |
| Data Rate | ~1 – 2 GB/day |
| Focal Length | 55 – 60 cm |
| Angular Resolution<br>Field of View (FoV) | 30 arcsec HPD<br>40 arcmin dia. |
| On-axis Effective Area<br>MiXO+CMOS+OBF Combined<br>(Module dependent: See Fig. 5) | ~$10^{-5}$ – $10^{-2}$ cm$^2$ at <1 keV<br>~$10^{-4}$ – $10^{-1}$ cm$^2$ at 1 – 2 keV<br>~$10^{-3}$ – ~1 cm$^2$ at 2 – 3 keV |
| Energy Resolution<br>Energy Range | <120 eV @ 1 keV<br>0.6 – 6 keV |
| Timing Resolution | ≲ 50 ms |

## 4.1 Miniature X-ray Optics (MiXO)

Miniature X-ray Optics (MiXO) are compact lightweight Wolter-I X-ray optics which are suitable for SmallSat missions. MiXO leverages the recent and on-going development to build lightweight Wolter-I X-ray optics based on the electroformed Ni-alloy replication (ENR) technique [e.g., 1, 34]. Replication consists of creating a substrate that faithfully reproduces the figure of a mandrel with high accuracy. Replication technology was used to fabricate the grazing incidence nested optics for *XMM-Newton* [35]. The individual shells of *XMM-Newton* were up to 1 mm thick, which was the state of the art for these large diameter, 15 arcsec optics at that time. More recently, X-ray optics with 25 and 20 arcsec angular resolution (HPD) have been fabricated for missions such as ART-XC [36] and *FOXSI* [37, 38] respectively, using shell thicknesses of 250 microns. And this number is improving as mounting techniques improve.

For the fabrication of ENR optics, thin nickel or nickel-alloy shells are electroformed onto figured and polished electroless-nickel-plated aluminum mandrels from which they are later separated by differential thermal contraction [39]. The surface and figure of the mandrel are replicated by the optic; good quality X-ray optics requiring both low (<5Å) surface microroughness and good (< 15 arcsec) mandrel figure. The attraction of the ENR process for thin-shell X-ray mirrors is that the resulting optics are full shells of revolution and this makes them inherently stable with good figure control, which offers the potential for good angular resolution. To date, nickel electroforming is the method that has produced the best replicated mirrors.

The effective area of grazing incidence optics is dictated both by the size of the optics and the number of nested shells. For flux-starved astrophysics missions, large effective area is achieved by nesting many dozens of shells (*NuSTAR*, *XMM-Newton*). However, for *SSAXI*, only a few nested shells (e.g., 1 – 2 shells for low and medium energy modules, ~5 – 10 shells for high energy modules) are required because of such high flux from the Sun. This reduces both the time and cost of fabricating the X-ray optics modules. And since several shells can be replicated from a single mandrel, additional mirrors modules can be fabricated at a relatively small additional cost.

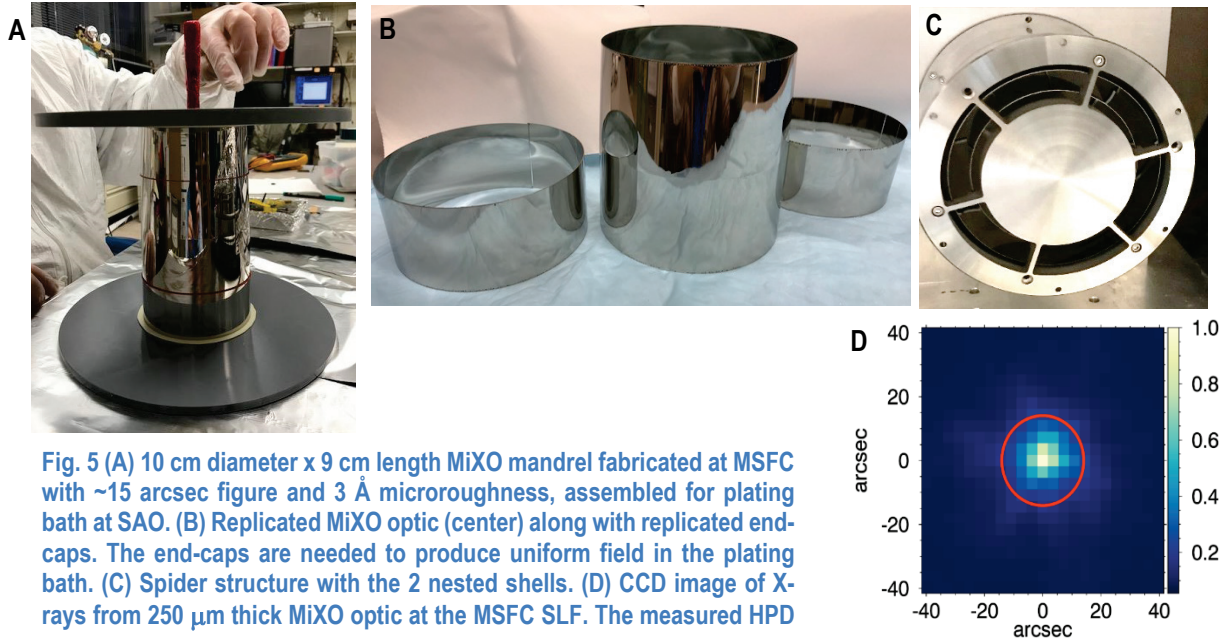

**Fig. 5 (A)** 10 cm diameter x 9 cm length MiXO mandrel fabricated at MSFC with ~15 arcsec figure and 3 Å microroughness, assembled for plating bath at SAO. **(B)** Replicated MiXO optic (center) along with replicated end-caps. The end-caps are needed to produce uniform field in the plating bath. **(C)** Spider structure with the 2 nested shells. **(D)** CCD image of X-rays from 250 μm thick MiXO optic at the MSFC SLF. The measured HPD

SAO in collaboration with MSFC and a few industry partners has multiple programs to develop replicated optics for a number of applications. Among these, Miniature X-ray Optics (MiXO) for planetary application is funded for development under a NASA PICASSO program. The primary objective of the PICASSO effort is to develop small and light weight ENR mirror shells suitable for planetary applications [40, 41].

A MiXO mandrel for the PICASSO program along with a replicated shell is shown in **Fig. 5**. Metrology of the mandrel indicates a figure of 15 arcsec. We expect the best shell replicated from this mandrel to have a resolution of ~ 23 arcsec (HPD) due to stresses introduced during the replication and separation process. A few of these replicated shells have been characterized at the Stray Light Facility (SLF) at MSFC and have a HPD of 30 arcsec, as shown in **Fig. 5D**.

**Fig. 6** illustrates the design concept for the MiXO in the high energy CXIS module for *SSAXI,* where the optics consists of 10 nested shells an optics housing and "spider"/spoke support structure. The individual shells will be 250 microns thick, with diameters ranging from ~ 4 – 7 cm and length of ~ 6 – 7 cm. This particular concept uses a butterfly design where the shell lengths vary with the shell diameter in order to allow the wide field over the wide energy range. For low and medium energy CXIS modules of *SSAXI*, the optics with one or two nested shells are expected to provide more than enough effective area (see **Section 5**) to achieve the Science Objectives.

### 4.2 CMOX X-ray sensor

CMOS imagers are well-suited to be the next generation of pixelated Si detectors for space-based optical, IR and X-ray telescopes due to their performance, operational simplicity, and inherent tolerance to radiation. The rapid readout makes CMOS attractive for *SSAXI*, allowing the coverage of a wide dynamic range of the solar flux. The focal plane in each *SSAXI* CXIS module employs a monolithic CMOS imager in a molybdenum (Mo) package. The cover of the Mo package incorporates a 40-arcminute wide circular aperture to define the science FoV. The main candidate CMOS Sensor for *SSAXI* is Big Minimal III (BM-III) developed and tested by SRI International in collaboration with SAO. **Fig. 7** shows a concept design of the *SSAXI* focal plane and **Fig. 8** shows the pictures of a BM-III mounted on a Headboard in a laboratory packaging. The BM-III devices share a common heritage with the flight CMOS imagers provided by SRI for other programs [42, 43] such as *SoloHi*, WISPR [44], and the Europa Imaging System (EIS) [45].

The pixel and signal chains of the BM-III devices were designed by SRI International (Sarnoff), following the performance specification from SAO. The layout of the devices was carried out by Chronicle Technology. JAZZ-Tower Semiconductor fabricated the device wafers through 180 nm process. Unlike the *SoloHI* and WISPR devices which are front illuminated, the X-ray optimized BM-III is back-thinned for Back-Illumination. The backside thinning and processing were performed at the University of Arizona Imaging Technology Lab (ITL) [46]. The ITL X-ray optimized backside process [44] has flight heritage as it was used on the JAXA mission *Suzaku* [47] X-ray CCDs. For *SSAXI*, front illuminated devices can be also used.

The BM-III device is comprised of a 1k × 1k array of 16 µm 6-Transistor Pinned Photo Diode (6T PPD) pixels over 1.6 × 1.6 cm. The devices are fabricated on epitaxial silicon and have a depletion depth of 10 µm. Each pixel has its own collection node, sense node, and source follower amplifier; hence charge to voltage conversion occurs at the pixel with a very high gain (high sensitivity) of ~135 µV/e⁻. The high sensitivity pixel enables the device to detect and resolve X-rays with energies well below 1 keV (e.g. for a carbon X-ray which produces on average 77 photo-electrons, the resulting voltage at the pixel would be >10mV). **Fig. 8** shows the spectral response of a BM-III device for Mg-K and O-K lines measured at the room temperature at the SAO. The 16 µm pixel sufficiently oversamples 30 arcsec HPD and the 40 arcmin FoV is covered by a quarter of the single device.

The maximum possible read rate of a full frame for BM-III is 40 Hz [46]. Smaller regions of interest can be read out at proportionally much higher rates. The *SSAXI* science FoV will be limited to a window of ~500 × 500 pixels, which can be read out at rates of up >~100 Hz. Unlike CCDs, CMOS pixels inherently incorporate anti-blooming so the rest of the detector not being addressed or "reset" will not affect operation or performance of the science sub-window. Since CMOS pixels are essentially randomly addressable, the size and location of the window can be easily changed.

The optical block filter for CMOS will vary depending on energy band and flux range coverage of *SSAXI* CXIS modules. Each CXIS module will cover either soft (0.5 – 1.2 keV), medium (1.5 – 2.5 keV) or hard (2.5 – 7 keV) energy range. The flux coverage will depend on the regions of Sun (quiet vs. active regions). The hard

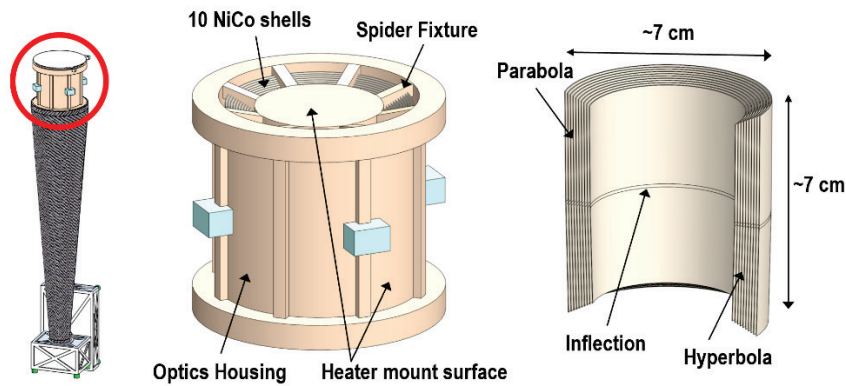

**Fig. 6 Example MiXO configuration for high energy CXIS module of *SSAXI*. This particular configuration is designed to achieve 30 – 40 arcsec HPD over 40 arcmin diameter FoV with ~ 1 – 2 cm² on-axis effective area in 3 – 4 keV band.**

X-ray modules will require thick OBF employing 400 – 1000 µm Be and medium X-ray modules will require medium size OBF of 100 – 200 µm Be to suppress soft X-rays from dominating readout signals.

A single common Controller Board (CTB) powers, controls, and processes the data from four CMOS imagers of *SSAXI*. The software used to process X-ray events detected by the CMOS X-ray sensor, is nearly identical to that of CCD sensors. A median map of several dark frames for the imager array is periodically generated and stored, then during nominal operation, the median frame is subtracted from X-ray imaging frames to eliminate any remaining Fixed Pattern Noise (FPN). The FPN corrected X-ray images are then processed to extract X-ray event location and energy. The energies of the 3 x 3 pixels around the triggered pixel are packaged and telemetered along with the pixel coordinates, telescope ID, and time tag. The CTB also collects and telemeters various housekeeping (HK) information from the temperature sensors, voltage and current monitors of each CXIS periodically.

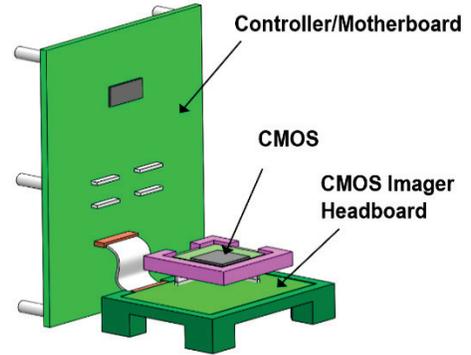

Fig. 7 Focal plane design of a CXIS module in *SSAXI* along with the main controller board

In order to limit the data rate to ~1 – 2 GB per day for each CXIS module while maintaining the full spectral and temporal information of the active region and microflares, a dynamic software filter that randomly drops X-ray events in active regions will be employed if needed. The random event filter enables the full spectral information of sufficient statistics in a reasonable data bandwidth. For modules optimized for axion search in the quiet Sun region, an alternative approach for efficient data transfer can be considered, where spatially divided daily spectral histograms (about 5000 spectra over entire solar disc) can be accumulated and downlinked (100 – 200 MB /day). Each spectrum is accumulated for a designated region of ~0.5 – 1 arcmin wide in the solar disc when the region does not exhibit any significant solar activities. A set of counters corresponding the collective counts of each region can be used to track the live activity of the region to properly select the quiet time intervals for histogram accumulation and normalization.

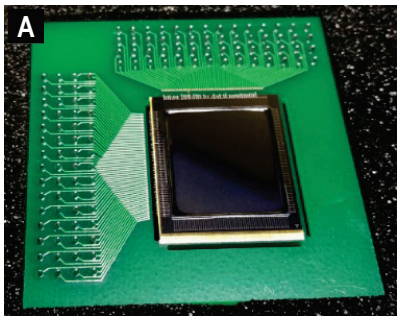 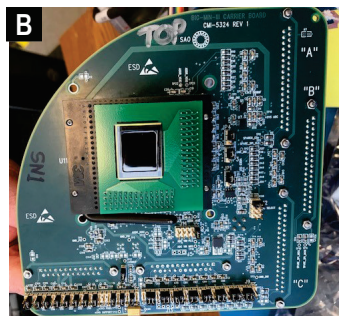 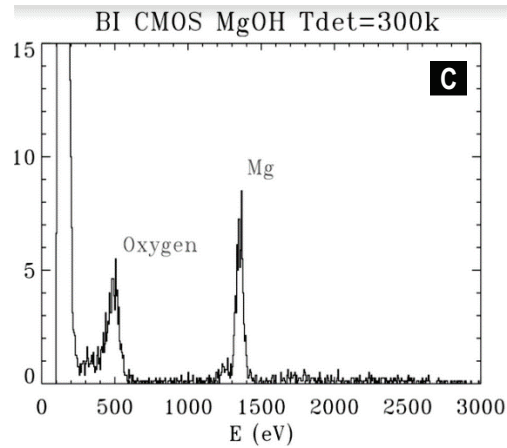

Fig. 8 (A) SAO/SRI BM III: 1k x 1k pixels, 16 µm pitch, Back Illuminated (BI). (B) BM III on a CMOS Headboard. (C) X-ray spectrum obtained with CMOS BM-III device. Figure shows Mg and Oxygen X-ray lines obtained with a Manson Model 5 electron impact source. Data were collected with the detector at T=300K.

## 5. *SSAXI* MISSION DESIGN

**Fig. 9** shows an example mission concept for *SSAXI* consisting of 6 CXIS modules and MicroSat S5 spacecraft from Blue Canyon Technologies (BCT). Given the wide dynamic range of the solar flux that is highly energy dependent, each CXIS module is optimized for specific energy and flux ranges, so that 3 modules are optimized for soft (0.5 – 1.2 keV), medium (1 – 2.5 keV), and hard X-rays (2 – 6 keV) from the Quiet Sun and the other 3 modules are for those from the Active Regions. These modules collectively can cover the broad 0.5 – 6 keV band for both quiet and active regions in the Sun.

*SSAXI* in **Fig. 9** is designed to rideshare into a LEO or more preferably into a Sun Synchronous orbit as a secondary payload of the ESPA-standard class. In the case of the Sun-Synchronous orbit, *SSAXI* can conduct nearly continuous monitoring of the entire solar disc for the planned 1 yr science operation at relatively stable thermal environments. Since the *SSAXI* instruments and the spacecraft do not require any consumables (e.g., the thermal system consists of passive cooling through radiators and active heating through trim heaters), it is expected that the science operation can be easily extended beyond the planned 1-yr term to maximize the science return. Solar observations will be periodically interrupted for Calibration Operations (e.g., the Q.E. measurements using the Crab Nebula). The focal plane will be equipped with onboard $^{55}$Fe calibration sources that enable continuous tracking of the spectral gain and resolution of the CMOS X-ray sensors and the subsequent readout system.

## 6. SUMMARY

We have introduced the new SmallSat mission concept, *SSAXI*, which is designed to search X-ray signature of solar axions and to study X-ray properties of microflares to understand the coronal heating processes. The CXIS, the main instrument of *SSAXI*, is a small Wolter-I X-ray telescope. Depending on the spacecraft capability, *SSAXI* can be packaged with multiple CXIS modules to enhance the science return. As an example, we have presented a *SSAXI* mission concept using a MicroSat S5 spacecraft available from BCT, which can carry and operate 6 CXIS modules for the ESPA-standard class payload. The *SSAXI* mission using MicroSat S5 rideshare to a LEO or Sun-Synchronous orbit for 1 yr science operation. With low-cost commercial spacecrafts, SmallSat missions like *SSAXI* can be developed in a short time scale.

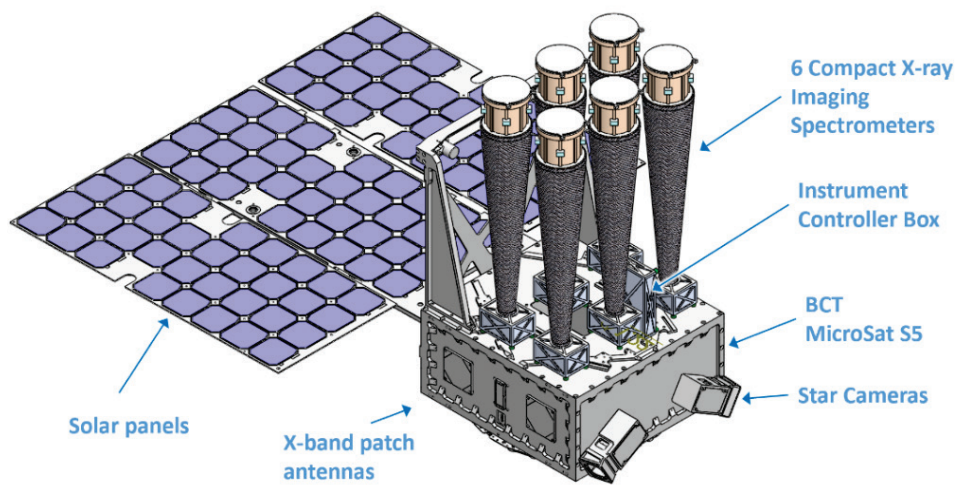

**Fig. 9** Example *SSAXI* mission concept consisting of 6 CXIS modules and BCT MicroSat S5 spacecraft. The 6 modules are divided into 3 energy bands (soft, medium, hard X-rays) x 2 flux ranges (Quiet Sun vs Active Regions).

## ACKNOWLEGEMENTS

The development of MiXO is supported by the NASA PICASSO program NNX16AL75G. Part of this work was performed under the auspices of the U.S. Department of Energy by Lawrence Livermore National Laboratory under Contract No. DE-AC52-07NA27344.